\documentclass[a4paper,twoside,12pt]{article}
\input{obsjkt.sty}

\newcommand{\Msun}{\ensuremath{~{\rm M}_\odot}}                   
\newcommand{\Rsun}{\ensuremath{~{\rm R}_\odot}}                   
\newcommand{\rhosun}{\ensuremath{~\rho_\odot}}                    
\newcommand{\Teff}{\ensuremath{T_{\rm eff}}}                      
\newcommand{\logg}{\ensuremath{\log g}}                           
\newcommand{\Vsys}{\ensuremath{V_\gamma}}                         
\newcommand{\EBV}{\ensuremath{E(B\!-\!V)}}                        
\newcommand{\Grp}{\ensuremath{G_{\rm RP}}}                        
\newcommand{\degr}{\ensuremath{^\circ}}                           
\renewcommand{\kms}{~km~s$^{-1}$}                                 
\newcommand{\chir}{\ensuremath{\chi_\nu^{\,2}}}                   
\newcommand{\gaia}{\textit{Gaia}}                                 
\newcommand{\targ}{UZ~Dra}
\newcommand{\targfull}{UZ~Draconis}

\newcommand{\Msunnom}{\hbox{$\mathcal{M}^{\rm N}_\odot$}}
\newcommand{\Rsunnom}{\hbox{$\mathcal{R}^{\rm N}_\odot$}}
\newcommand{\Lsunnom}{\hbox{$\mathcal{L}^{\rm N}_\odot$}}

\usepackage{rotating}

\begin{document} 

\OBSheader{Rediscussion of eclipsing binaries: \targ}{J.\ Southworth}{2025 December}

\OBStitle{Rediscussion of eclipsing binaries. Paper XXVII. \\ The totally-eclipsing system UZ~Draconis}

\OBSauth{John Southworth}

\OBSinstone{Astrophysics Group, Keele University, Staffordshire, ST5 5BG, UK}


\OBSabstract{\targ\ is a detached and totally-eclipsing binary containing two late-F stars in a circular orbit of period 3.261~d. It has been observed by the Transiting Exoplanet Survey Satellite in 41 sectors, yielding a total of 664,809 high-quality flux measurements. We model these data and published radial velocities to determine the physical properties of the system to high precision. The masses of the stars are $1.291 \pm 0.012$\Msun\ and $1.193 \pm 0.009$\Msun, and their radii are $1.278 \pm 0.004$\Rsun\ and $1.122 \pm 0.003$\Rsun. The high precision of the radius measurements is made possible by the (previously unrecorded) total eclipses and the extraordinary amount of data available. The light curves show spot modulation at the orbital period, and both stars rotate synchronously. Our determination of the distance to the system, $185.7 \pm 2.4$~pc, agrees very well with the parallax distance of $185.39 \pm 0.39$~pc from \gaia\ DR3. The properties of the system are consistent with theoretical predictions for an age of $600 \pm 200$~Myr and a slightly super-solar metallicity.}


\section*{Introduction}

Detached eclipsing binaries (dEBs) are our primary source of direct measurements of the basic physical properties of normal stars \cite{Andersen91aarv,Torres++10aarv,Me15aspc} because their masses and radii can be determined from light and radial velocity (RV) curves using only geometry and celestial mechanics. Within this class of object, those that have total eclipses are the most valuable because the times of contact during eclipse enable the radii of the stars to be measured to the highest precision \cite{Russell12apj,Maxted+20mn}.

In this work we present an analysis of the late-F-type dEB \targ, which shows total eclipses and a circular orbit. This analysis is part of our project to systematically redetermine the properties of known dEBs using new space-based light curves \cite{Me21univ} and published spectroscopic results \cite{Me20obs}.

\section*{\targfull}

\begin{table}[t]
\caption{\em Basic information on \targfull. 
The $BV$ magnitudes are each the mean of 115 individual measurements \cite{Hog+00aa} distributed approximately randomly in orbital phase. 
The $JHK_s$ magnitudes are from 2MASS \cite{Cutri+03book} and were obtained at an orbital phase of 0.296. \label{tab:info}}
\centering
\begin{tabular}{lll}
{\em Property}                            & {\em Value}                 & {\em Reference}                      \\[3pt]
Right ascension (J2000)                   & 19 25 55.054                & \citenum{Gaia23aa}                   \\
Declination (J2000)                       & +68 56 07.16                & \citenum{Gaia23aa}                   \\
\textit{Tycho} designation                & TYC 4444-1595-1             & \citenum{Hog+00aa}                   \\
\textit{Gaia} DR3 designation             & 2261658485914111744         & \citenum{Gaia21aa}                   \\
\textit{Gaia} DR3 parallax (mas)          & $5.3941 \pm 0.0115$         & \citenum{Gaia21aa}                   \\          
TESS\ Input Catalog designation           & TIC 48356677                & \citenum{Stassun+19aj}               \\
$B$ magnitude                             & $10.08 \pm 0.03$            & \citenum{Hog+00aa}                   \\          
$V$ magnitude                             & $9.60 \pm 0.02$             & \citenum{Hog+00aa}                   \\          
$J$ magnitude                             & $8.616 \pm 0.020$           & \citenum{Cutri+03book}               \\
$H$ magnitude                             & $8.426 \pm 0.020$           & \citenum{Cutri+03book}               \\
$K_s$ magnitude                           & $8.372 \pm 0.019$           & \citenum{Cutri+03book}               \\
Spectral type                             & F6 + F8                     & \citenum{Popper96apjs}               \\[3pt]
\end{tabular}
\end{table}



The variability of \targ\ (Table~\ref{tab:info}) was announced by Pickering \cite{Pickering07aj}, following its discovery by Henrietta Leavitt in photographic patrol plates from Harvard. It was awarded the designation `HV 2972', its range of variation was given as 0.7~mag, and its variability type was described using the phrase ``appear[s] to be of the Algol type''. 

Dugan \& Wright \cite{DuganWright37aj} found an orbital period of 1.63~d, half the true period because the secondary eclipses were mistaken as primaries. Lacy et al.\ \cite{Lacy+89aj} (hereafter L89) state that a doubled period of 3.26~d was adopted by Tsesevitch \cite{Tsesevitch54izaoo}. This was confirmed and refined by Koch \& Koch \cite{KochKoch62aj} using brightness measurements from 35~mm film. G\"ulmen et al.\ \cite{Gulmen++86ibvs} collected all times of minimum up to the year 1986.

Imbert \cite{Imbert86aas} presented the first spectroscopic orbits of \targ, obtaining precise velocity amplitudes ($K_{\rm A}$ and $K_{\rm B}$) from 40 RVs per star measured with the CORAVEL cross-correlation spectrometer \cite{Baranne++79va}. Lacy \cite{Lacy84ibvs} found it to be a double-lined binary system.

L89 presented the first -- and so far only -- detailed study of \targ. This was based on 35 nights of photoelectric $BV$ photometry from Ege University (ref.~\citenum{Gulmen++86ibvs}) and 16 high-resolution spectra from two telescopes. Six spectra were obtained with the coud\'e spectrograph and Reticon detector on the 2.7~m telescope at McDonald Observatory, and the remaining ten with the coud\'e spectrograph and a CCD detector on the 2.1~m telescope at Kitt Peak National Observatory. From analysis of these material they measured the masses and radii of the component stars to precisions of 1.5--2.3\%. They also obtained projected rotational velocities of $20 \pm 1$\kms\ and $19 \pm 1$\kms, both consistent with synchronous rotation in the assumed circular orbit, spectral types of F7 and G0, and a spectroscopic light ratio of $0.73 \pm 0.03$ from the 6400~\AA\ Fe~I and 6439.1~\AA\ Ca~I lines.

Since that work, Popper \cite{Popper96apjs} has indicated spectral types of F6 and F8 for the two stars, and Graczyk et al.\ \cite{Graczyk+17apj} have presented updated masses, radii and temperatures of the stars. A large number of times of eclipse are also available; \targ\ is a popular target for amateur astronomers.


\section*{Photometric observations}

\targ\ has been observed by the NASA Transiting Exoplanet Survey Satellite \cite{Ricker+15jatis} (TESS) in an extraordinary 41 sectors to date, due to its placement within the satellite's northern continuous viewing zone. In all cases data are available at 120~s cadence from the SPOC (Science Processing Center \cite{Jenkins+16spie}). Lower-cadence observations are also available for all sectors but were not used here. The data were downloaded from the NASA Mikulski Archive for Space Telescopes (MAST\footnote{\texttt{https://mast.stsci.edu/portal/Mashup/Clients/Mast/Portal.html}}) using the {\sc lightkurve} package \cite{Lightkurve18}. 

We used the simple aperture photometry (SAP) light curves from the SPOC data reduction pipeline \cite{Jenkins+16spie} for our analysis, and rejected low-quality data using the {\sc lightkurve} quality flag ``hard''.  A total of 664,809 datapoints survived this cut, coming from TESS sectors 14 to 86. These data were converted into differential magnitude and the median magnitude was subtracted from each sector for convenience. 

\begin{figure}[t] \centering \includegraphics[width=\textwidth]{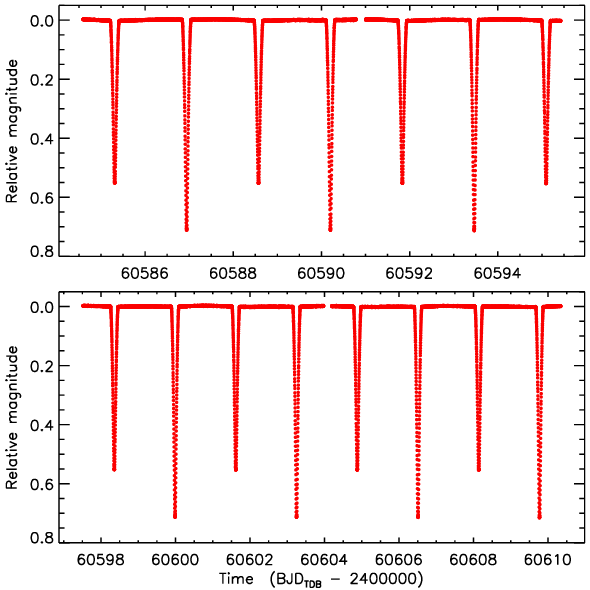} \\
\caption{\label{fig:time} TESS sector 84 photometry of \targ. The flux measurements  
have been converted to magnitude units after which the median was subtracted. The   
other sectors used in this work are very similar so are not plotted.} \end{figure}

Fig.~\ref{fig:time} shows the light curve from sector 84, chosen because of its high duty cycle; the remaining sectors are similar so are not plotted. One feature of the light curve immediately apparent on closer inspection is that the eclipses are total. This seems not to have been noticed previously, which led L89 to use a spectroscopic light ratio to constrain the ratio of the radii of the stars measured from the eclipse shapes.

We queried the \gaia\ DR3 database\footnote{\texttt{https://vizier.cds.unistra.fr/viz-bin/VizieR-3?-source=I/355/gaiadr3}} for all sources within 2~arcmin of \targ. Of the sources returned -- 53 excluding the dEB itself -- all are at least 5~mag fainter in the \gaia\ \Grp\ band so should contribute little additional flux to the TESS light curves.


\section*{Light curve analysis}

The components of \targ\ are well-separated so we modelled the TESS light curves using the version 44 of the {\sc jktebop}\footnote{\texttt{http://www.astro.keele.ac.uk/jkt/codes/jktebop.html}} code \cite{Me++04mn2,Me13aa}. We defined star~A to be the star eclipsed at the primary (deeper) minimum, and star~B to be its companion. Star~A is hotter, larger and more massive than star~B.

The exceptional amount of data necessitated the analysis of each sector separately, which in turn required the automation of some tasks usually performed manually. For each TESS sector we chose a primary eclipse close to the midpoint of the light curve, rejected data with large scatter or close to partially-observed eclipses, and defined normalisation polynomials to remove slow variations in the measured brightness of the system. A total of 608,575 datapoints were retained for analysis. 

\begin{figure}[t] \centering \includegraphics[width=\textwidth]{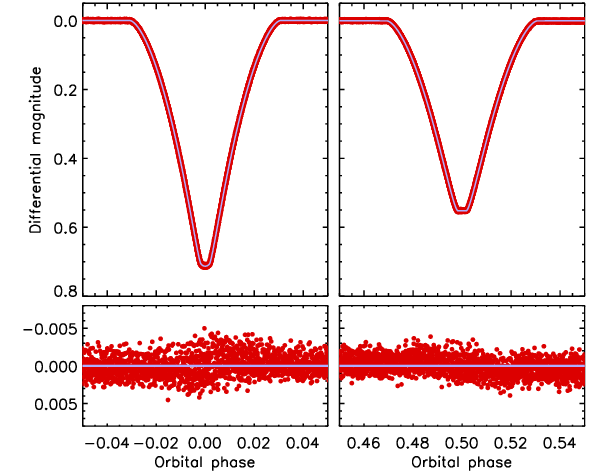} \\
\caption{\label{fig:phase} {\sc jktebop} best fit to the light curves of \targ\ from 
TESS sector 84 for the primary eclipse (left panels) and secondary eclipse (right panels). 
The data are shown as filled red circles and the best fit as a light blue solid line. 
The residuals are shown on an enlarged scale in the lower panels.} \end{figure}

We then fitted the data from each TESS sector using {\sc jktebop} with the following fitted parameters: the fractional radii of the stars ($r_{\rm A}$ and $r_{\rm B}$) taken as the sum ($r_{\rm A}+r_{\rm B}$) and ratio ($k = r_{\rm B}/r_{\rm A}$), the central surface brightness ratio ($J$), third light ($L_3$), orbital inclination ($i$), orbital period ($P$), and a reference time of primary minimum ($T_0$). A circular orbit provides a good fit to all data so we assumed an eccentricity of zero; experiments with a fitted eccentricity caused changes in the measured fractional radii at approximately the 0.004\% level. Limb darkening (LD) was accounted for using the power-2 law \cite{Hestroffer97aa,Maxted18aa,Me23obs2}, the linear coefficients ($c$) were fitted, and the non-linear coefficients ($\alpha$) were fixed at theoretical values \cite{ClaretSouthworth22aa,ClaretSouthworth23aa}. The measurement errors were scaled to force a reduced $\chi^2$ of $\chir = 1.0$. An example fit is shown in Fig.~\ref{fig:phase}.

\begin{table} \centering
\caption{\em \label{tab:jktebop} Photometric parameters of \targ\ measured using 
{\sc jktebop} from the light curves from all 41 TESS sectors. The errorbars are 
standard deviations (not standard errors) of the results for individual sectors.}
\begin{tabular}{lcc}
{\em Parameter}                           &              {\em Value}            \\[3pt]
{\it Fitted parameters:} \\
Orbital inclination (\degr)               & $      89.708      \pm  0.024     $ \\
Sum of the fractional radii               & $       0.19154    \pm  0.00009   $ \\
Ratio of the radii                        & $       0.87795    \pm  0.00076   $ \\
Central surface brightness ratio          & $       0.8582     \pm  0.0051    $ \\
Third light                               & $       0.0104     \pm  0.0045    $ \\
LD coefficient $c_{\rm A}$                & $       0.611      \pm  0.013     $ \\
LD coefficient $c_{\rm B}$                & $       0.568      \pm  0.025     $ \\
LD coefficient $\alpha_{\rm A}$           &              0.4984 (fixed)         \\
LD coefficient $\alpha_{\rm B}$           &              0.5237 (fixed)         \\
{\it Derived parameters:} \\
Fractional radius of star~A               & $       0.10199    \pm  0.00007   $ \\       
Fractional radius of star~B               & $       0.08955    \pm  0.00006   $ \\       
Light ratio $\ell_{\rm B}/\ell_{\rm A}$   & $       0.6641     \pm  0.0025    $ \\[3pt]
\end{tabular}
\end{table}

Table~\ref{tab:jktebop} lists the results of this analysis. For each parameter we took the final value and errorbar to be the unweighted mean and standard deviation of the values from the individual sectors. We did not convert the standard deviation into a standard error because it is already at the limit to which we trust our photometric model for some parameters -- in particular the fractional radii in {\sc jktebop} have been shown to be reliable to 0.1\% precision \cite{Maxted+20mn} but not beyond.

We also calculated uncertainties using Monte Carlo simulations\footnote{Running 500 Monte Carlo simulations for each light curve required approximately 40 hours of computing time on a standard-specification laptop.} to provide errorbars on all parameters. The Monte Carlo errorbars are smaller than the standard deviation of the values by factors of 1--3 in the case of \targ, likely due to the influence of spot modulation on the light curve fits (see below)

Fig.~\ref{fig:parget} shows the variation of the most important photometric parameters with time, with one datapoint for each TESS sector. The small variation in the parameters is striking, confirming the reliability of the solutions for \targ. No significant slow variations with time are apparent. The Monte Carlo errorbars underestimate the true uncertainty in the light ratio, an issue which may be caused by the spot modulation in the light curve. The third light is not expected to be the same between sectors due to the different pixel position of \targ\ and pixel mask used each time the telescope is re-orientated.

\begin{figure}[t] \centering \includegraphics[width=\textwidth]{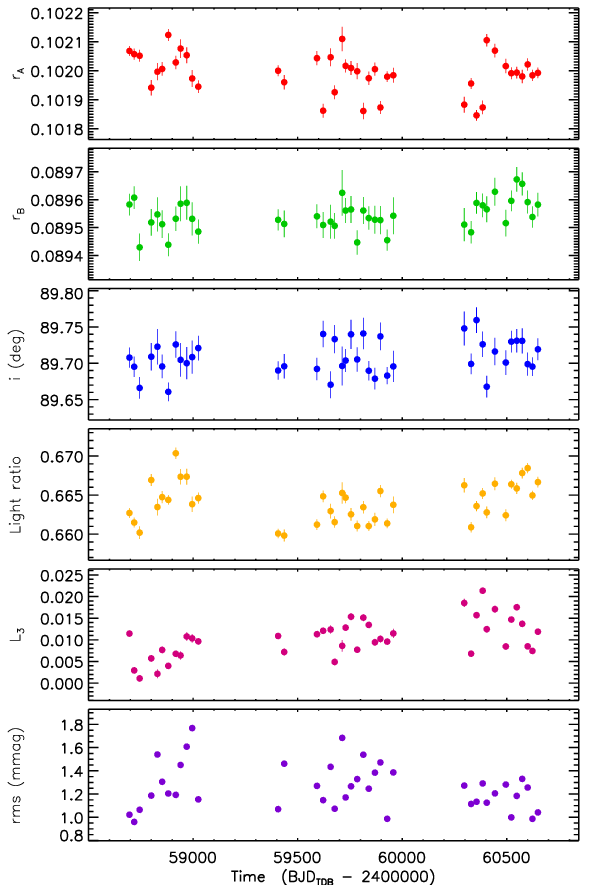} \\
\caption{\label{fig:parget} The best fit to selected photometric parameters of
\targ\ from TESS sectors 1 to 86. The times used in the plot are those presented
in the next section. The errorbars are from Monte Carlo simulations.} \end{figure}


\section*{Orbital ephemeris}

\begin{table} \centering
\caption{\em Times of mid-eclipse for \targ\ and their residuals versus the fitted ephemeris. \label{tab:tmin}}
\setlength{\tabcolsep}{10pt}
\begin{tabular}{rrrrr}
{\em Orbital} & {\em Eclipse time}  & {\em Uncertainty} & {\em Residual} & {\em TESS}   \\
{\em cycle}   & {\em (BJD$_{TDB}$)} & {\em (d)}         & {\em (d)}      & {\em sector} \\[3pt]
$-301.0$ & 2458695.386152 & 0.000004 & $ $0.000001 & 14 \\   
$-294.0$ & 2458718.215246 & 0.000005 & $-$0.000026 & 15 \\   
$-286.0$ & 2458744.305680 & 0.000006 & $-$0.000016 & 16 \\   
$-269.0$ & 2458799.747840 & 0.000006 & $-$0.000008 & 18 \\   
$-260.0$ & 2458829.099615 & 0.000006 & $ $0.000040 & 19 \\   
$-253.0$ & 2458851.928689 & 0.000006 & $-$0.000007 & 20 \\   
$-244.0$ & 2458881.280400 & 0.000005 & $-$0.000024 & 21 \\   
$-233.0$ & 2458917.154773 & 0.000005 & $ $0.000016 & 22 \\   
$-226.0$ & 2458939.983865 & 0.000007 & $-$0.000013 & 23 \\   
$-217.0$ & 2458969.335652 & 0.000006 & $ $0.000046 & 24 \\   
$-209.0$ & 2458995.426050 & 0.000007 & $ $0.000020 & 25 \\   
$-200.0$ & 2459024.777740 & 0.000005 & $-$0.000017 & 26 \\   
$ -83.0$ & 2459406.350163 & 0.000004 & $-$0.000050 & 40 \\   
$ -74.0$ & 2459435.701982 & 0.000005 & $ $0.000042 & 41 \\   
$ -26.0$ & 2459592.244467 & 0.000005 & $-$0.000019 & 47 \\   
$ -17.0$ & 2459621.596224 & 0.000005 & $ $0.000011 & 48 \\   
$  -6.0$ & 2459657.470594 & 0.000007 & $ $0.000048 & 49 \\   
$   0.0$ & 2459677.038369 & 0.000005 & $ $0.000004 & 50 \\   
$  11.0$ & 2459712.912675 & 0.000011 & $-$0.000023 & 51 \\   
$  16.0$ & 2459729.219231 & 0.000005 & $ $0.000018 & 52 \\   
$  24.0$ & 2459755.309623 & 0.000006 & $-$0.000014 & 53 \\   
$  33.0$ & 2459784.661343 & 0.000005 & $-$0.000022 & 54 \\   
$  42.0$ & 2459814.013130 & 0.000006 & $ $0.000038 & 55 \\   
$  50.0$ & 2459840.103558 & 0.000004 & $ $0.000042 & 56 \\   
$  59.0$ & 2459869.455257 & 0.000005 & $ $0.000013 & 57 \\   
$  67.0$ & 2459895.545631 & 0.000005 & $-$0.000037 & 58 \\   
$  77.0$ & 2459928.158697 & 0.000004 & $-$0.000001 & 59 \\   
$  86.0$ & 2459957.510395 & 0.000008 & $-$0.000031 & 60 \\   
$ 190.0$ & 2460296.685987 & 0.000007 & $ $0.000045 & 73 \\   
$ 200.0$ & 2460329.298995 & 0.000004 & $ $0.000023 & 74 \\   
$ 208.0$ & 2460355.389387 & 0.000004 & $-$0.000009 & 75 \\   
$ 217.0$ & 2460384.741110 & 0.000005 & $-$0.000014 & 76 \\   
$ 223.0$ & 2460404.308962 & 0.000005 & $ $0.000020 & 77 \\   
$ 235.0$ & 2460443.444577 & 0.000005 & $-$0.000001 & 78 \\   
$ 251.0$ & 2460495.625437 & 0.000005 & $ $0.000010 & 80 \\   
$ 259.0$ & 2460521.715823 & 0.000004 & $-$0.000028 & 81 \\   
$ 267.0$ & 2460547.806279 & 0.000004 & $ $0.000004 & 82 \\   
$ 275.0$ & 2460573.896693 & 0.000005 & $-$0.000007 & 83 \\   
$ 283.0$ & 2460599.987162 & 0.000005 & $ $0.000038 & 84 \\   
$ 290.0$ & 2460622.816217 & 0.000004 & $-$0.000028 & 85 \\   
$ 298.0$ & 2460648.906647 & 0.000005 & $-$0.000022 & 86 \\   
\end{tabular}
\end{table}


Our photometric analysis above yielded a measurement of the mean time of primary eclipse for each TESS sector. We fitted a linear ephemeris to these times, obtaining
\begin{equation}
\mbox{Min~I} = {\rm BJD}_{\rm TDB}~ 2459677.038365 (4) + 3.261303037 (20) E
\end{equation}
in the barycentric rest frame, where $E$ is the number of cycles since the reference time of minimum and the bracketed quantities indicate the uncertainty in the final digit of the previous number. The scatter around the best fit is larger than the errorbars suggest, with $\chir = 25.4$, likely due to the weak spot activity visible outside eclipse in most TESS sectors. The uncertainties in the ephemeris have been multiplied by $\sqrt{\chir}$ to account for this. The individual timings are given in Table~\ref{tab:tmin}.

The deep eclipses combined with the high quality of the available data yield a very precise ephemeris: the r.m.s.\ scatter around the best fit is only 2.2~s, and the period is measured to within $\pm$2~ms. We extrapolated it back to the ephemeris given by G\"ulmen et al.\ \cite{Gulmen++86ibvs} (HJD 2446227.4238) and found that it matched to within 18~s, after correcting the HJD to BJD and converting to the TDB timescale \cite{Eastman++10pasp}. Based on this and our timings, we see no evidence for nonlinearity in the orbital ephemeris. A more robust approach would require assembling the many published times of minimum for \targ, which is beyond the scope of the current work.



\clearpage

\section*{Radial velocity analysis}

\begin{figure}[t] \centering \includegraphics[width=\textwidth]{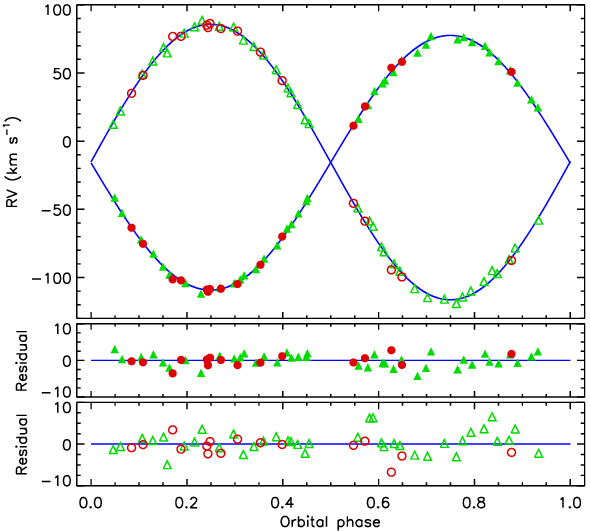} \\
\caption{\label{fig:rv} RVs of \targ\ compared to the best fit from 
{\sc jktebop} (solid blue lines). The RVs for star~A are shown with filled symbols, 
and for star~B with open symbols. The residuals are given in the lower panels 
separately for the two components. RVs from Imbert \cite{Imbert86aas} are shown 
with green triangles, and those from L89 with dark red circles.} \end{figure}

Two prior spectroscopic studies of \targ\ exist: Imbert \cite{Imbert86aas} and L89. The former obtained and analysed 40 RVs for star~A and 39 for star~B, using the CORAVEL spectrometer. The latter found that Imbert's results were slightly but significantly discrepant with their own measurements, based on 16 high-resolution spectra. This claimed disagreement is sufficient justification for us to revisit the RVs, both sets of which are tabulated in the respective papers.

\begin{sidewaystable} \centering
\caption{\em Spectroscopic orbits for \targ\ from the literature and from the current work. 
In each case two sets of orbits are given: where the systemic velocity for the two stars are 
forced to be the same or allowed to differ. The adopted result is based on all RVs and 
different systemic velocities. All quantities are given in km~s$^{-1}$. \label{tab:sb}}
\setlength{\tabcolsep}{10pt}
\begin{tabular}{lccccccc}
{\em Source} & {\em $K_{\rm A}$} & {\em $K_{\rm B}$} & {\em $V_{\rm \gamma}$} & {\em $V_{\rm \gamma,A}$} & {\em $V_{\rm \gamma,B}$} & {\em  $\sigma_{\rm A}$} & {\em $\sigma_{\rm B}$} \\[10pt]
Imbert \cite{Imbert86aas} 	          & $92.73 \pm 0.46$ & $100.51 \pm 0.46$ & $-15.29 \pm 0.25$ &                   &                   & 1.6~ & 2.5~ \\
L89							          & $94.3  \pm 0.5 $ & $102.6  \pm 0.7 $ &                   & $-15.6  \pm 0.4 $ & $-16.8  \pm 0.6 $ & 1.4~ & 2.0~ \\[10pt]
This work (Imbert RVs)		          & $92.87 \pm 0.34$ & $100.50 \pm 0.54$ & $-15.45 \pm 0.22$ &                   &                   & 1.63 & 2.57 \\
This work (Imbert RVs)		          & $92.87 \pm 0.36$ & $100.52 \pm 0.55$ &                   & $-15.71 \pm 0.26$ & $-14.80 \pm 0.40$ & 1.61 & 2.48 \\[10pt]
This work (L89 RVs)		              & $93.96 \pm 0.46$ & $102.05 \pm 0.64$ & $-15.99 \pm 0.31$ &                   &                   & 1.30 & 1.97 \\
This work (L89 RVs)		              & $94.02 \pm 0.48$ & $102.55 \pm 0.68$ &                   & $-15.64 \pm 0.37$ & $-16.73 \pm 0.56$ & 1.28 & 1.87 \\[10pt]
This work (all RVs)		              & $93.43 \pm 0.28$ & $101.13 \pm 0.43$ & $-15.69 \pm 0.20$ &                   &                   & 1.61 & 2.55 \\
\textbf{This work (all RVs, adopted)} & $93.40 \pm 0.30$ & $101.03 \pm 0.44$ &                   & $-15.84 \pm 0.21$ & $-15.33 \pm 0.34$ & 1.60 & 2.50 \\
\end{tabular}
\end{sidewaystable} 

We first digitised the Imbert RVs then performed two fits with {\sc jktebop}, one with the same systemic velocity for both stars (\Vsys) and one with a systemic velocity per star ($V_{\rm \gamma,A}$ and $V_{\rm \gamma,B}$). In both cases we fitted for $K_{\rm A}$, $K_{\rm B}$ and an offset from our ephemeris above (which turns out to be negligible), and assumed a circular orbit. All errorbars were obtained using 1000 Monte Carlo simulations \cite{Me21obs5}; the errorbars in the systemic velocities do not account for any systematic bias due to transformation onto a standard system. The two solutions are very similar (Table~\ref{tab:sb}), and confirm the numbers presented by Imbert \cite{Imbert86aas}.

We then undertook the same analysis for the RVs from L89. An identical picture emerged: a good agreement between our two solutions and the values given by L89. The phase offset was also negligible, a relevant point because the timestamps of the RVs are tabulated to only three decimal places (a precision of 86~s) in L89. We therefore confirmed the small but significant discrepancy between the two sets of measurements.

Faced with this choice, L89 opted to use only their own RVs on the basis that previous work by these authors had given results in good agreement with independent measurements for several dEBs. We have used RVs from Claud Lacy several times in the current series of papers: for ZZ~UMa \cite{LacySabby99ibvs,Me22obs6}, IT~Cas \cite{Lacy+97aj,Me23obs3}, IQ~Per \cite{LacyFrueh85apj,Me24obs5} and MU~Cas \cite{Lacy++04aj2,Me25obs1}. Similarly, we have in the past been happy to adopt RVs from Imbert for AN~Cam \cite{Imbert87aas,Me21obs3} and ZZ~UMa \cite{Imbert02aa,Me22obs6}; the two sources agreed well in the case of ZZ~UMa. 

We have therefore chosen to adopt spectroscopic orbits from the combined RVs for the remainder of our analysis. Neither source gives uncertainties for their RVs, so we specified uncertainties that give $\chir = 1.0$ for each of the four data sets (two per star). Our adopted fit has separate systemic velocities for the two stars, is shown in Fig.~\ref{fig:rv}, and its parameters are given in Table~\ref{tab:sb}. It is, unsurprisingly, intermediate between the spectroscopic orbits from Imbert \cite{Imbert86aas} and L89.


\section*{Physical properties and distance to \targ}

\begin{table} \centering
\caption{\em Physical properties of \targ\ defined using the nominal solar units 
given by IAU 2015 Resolution B3 (ref.~\citenum{Prsa+16aj}). \label{tab:absdim}}
\begin{tabular}{lr@{~$\pm$~}lr@{~$\pm$~}l}
{\em Parameter}        & \multicolumn{2}{c}{\em Star A} & \multicolumn{2}{c}{\em Star B}    \\[3pt]
Mass ratio   $M_{\rm B}/M_{\rm A}$          & \multicolumn{4}{c}{$0.9244 \pm 0.0050$}       \\
Semimajor axis of relative orbit (\Rsunnom) & \multicolumn{4}{c}{$12.533 \pm 0.034$}        \\
Mass (\Msunnom)                             &  1.291  & 0.012       &  1.193  & 0.009       \\
Radius (\Rsunnom)                           &  1.2783 & 0.0036      &  1.1224 & 0.0032      \\
Surface gravity ($\log$[cgs])               &  4.3356 & 0.0020      &  4.4145 & 0.0015      \\
Density ($\!\!$\rhosun)                     &  0.6178 & 0.0020      &  0.8438 & 0.0029      \\
Synchronous rotational velocity ($\!\!$\kms)& 19.830  & 0.056       & 17.411  & 0.049       \\
Effective temperature (K)                   &  6450   & 120         &  6170   & 120         \\
Luminosity $\log(L/\Lsunnom)$               &  0.406  & 0.032       &  0.216  & 0.033       \\
$M_{\rm bol}$ (mag)                         &  3.724  & 0.081       &  4.200  & 0.085       \\
Interstellar reddening \EBV\ (mag)          & \multicolumn{4}{c}{$0.02 \pm 0.01$}			\\
Distance (pc)                               & \multicolumn{4}{c}{$185.7 \pm 2.4$}           \\[3pt]
\end{tabular}
\end{table}


We calculated the physical properties of \targ\ using the {\sc jktabsdim} code \cite{Me++05aa} with the photometric properties from Table~\ref{tab:jktebop}, and the $K_{\rm A}$ and $K_{\rm B}$ from the previous section. The orbital period was corrected to the rest frame of the system using a systemic velocity of $-15.5$\kms. The masses are measured to 0.9\% precision, and the radii to 0.3\%. The radii agree to within approximately 1$\sigma$ with the measurements from L89, but the masses are slightly lower due to the choices made in the RV analysis. We adopted the effective temperatures of the stars of $6450 \pm 120$~K and $6170 \pm 120$~K, from Graczyk et al.\ \cite{Graczyk+17apj}. These are significantly higher than the $6200 \pm 120$ and $5985 \pm 110$~K given by L89 (see below for justification). For both sets of temperatures, their ratio is in good agreement with the central surface brightness ratio in Table~\ref{tab:jktebop}. 

The synchronous rotational velocities in Table~\ref{tab:absdim} are consistent with the projected rotational velocities measured by L89. Inspection of the residuals of the {\sc jktebop} fits shows that the spot modulation occurs on the orbital period of the system (see next section). We conclude that the system is tidally circularised and rotationally synchronised. There is no evidence or pulsations in the available data.

We determined the distance to \targ\ using the $BV$ magnitudes from Tycho \cite{Hog+00aa} and $JHK_s$ magnitudes from 2MASS \cite{Cutri+03book} given in Table~\ref{tab:info}. We used the surface brightness calibrations from Kervella et al.\ \cite{Kervella+04aa} and found that an interstellar reddening of $\EBV = 0.02 \pm 0.01$~mag was needed to bring the $BV$-based distance measurements into agreement with those from $JHK_s$. The best distance estimate is $185.7 \pm 2.4$~pc, which is in excellent agreement with the $185.39 \pm 0.39$~pc from the \gaia\ DR3 parallax \cite{Gaia23aa}. In contract, the temperatures from L89 give a shorter distance and need a slightly negative interstellar reddening to equalise the optical and IR distances.
 
We compared the measured masses, radii and temperatures of the stars to theoretical predictions from the {\sc parsec} 1.2 evolutionary models \cite{Bressan+12mn}. The best match occurs for an age of $600 \pm 200$~Myr and a slightly super-solar metal abundance of $Z=0.020$. Thus \targ\ is a relatively young dEB.


\section*{Stellar activity}



\begin{figure}[t] \centering \includegraphics[width=\textwidth]{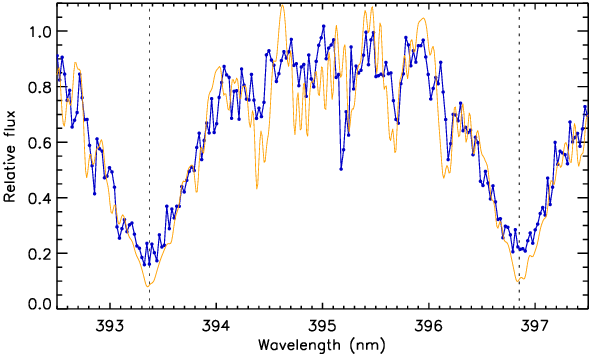} \\
\caption{\label{fig:cahk} Comparison between the observed spectrum of \targ\ and a 
synthetic spectrum around the Ca~{\sc ii} H and K lines. The observed spectrum has 
been shifted to zero velocity and normalised to approximately unit flux (blue line 
with points). The synthetic spectrum is for $\Teff = 6450$~K, $\logg = 4.3$ and solar 
metallicity from the BT-Settl model atmospheres \cite{Allard+01apj,Allard++12rspta} 
(orange line). A composite synthetic spectrum has been made by duplicating and 
shifting the original one to the velocities of the two components of \targ\ then 
adding them using the light ratio from Table~\ref{tab:jktebop}.} \end{figure}

We obtained a spectrum of the Ca~{\sc ii} H and K lines of \targ\ to search for evidence of emission caused by chromospheric activity, with the Intermediate Dispersion Spectrograph (IDS) at the Cassegrain focus of the Isaac Newton Telescope (INT). A single observation with an exposure time of 500~s was obtained on the night of 2022/06/07 in excellent weather conditions. We used the 235~mm camera, H2400B grating, EEV10 CCD and a 1~arcsec slit and obtained a resolution of approximately 0.05~nm. A central wavelength of 4050\,\AA\ yielded a spectrum covering 373--438~nm at a reciprocal dispersion of 0.023~nm~px$^{-1}$. The data were reduced using a pipeline currently being written by the author \cite{Me+20mn2}, which performs bias subtraction, division by a flat-field from a tungsten lamp, aperture extraction, and wavelength calibration using copper-argon and copper-neon arc lamp spectra.

The spectrum (Fig.~\ref{fig:cahk}) was obtained at orbital phase 0.887, when the RV difference of the two stars was 126\kms\ (0.17~nm). When compared to a composite synthetic spectrum without chromospheric activity \cite{Allard+01apj,Allard++12rspta}, the infilling of the H and K lines is clear. The velocity difference of the two stars is resolved and both show chromospheric emission, most obviously in the K line at 393.4~nm. 

\begin{figure}[t] \centering \includegraphics[width=\textwidth]{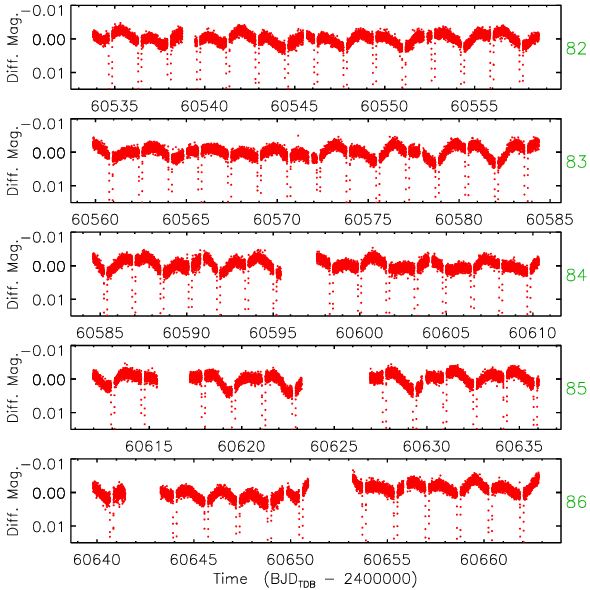} \\
\caption{\label{fig:spot} Differential-magnitude light curve of \targ\ from 
sectors 82 to 86 (labelled on the right of each panel). The y-axis has been 
chosen so the out-of-eclipse variation due to starspots is clear.} \end{figure}

The TESS light curves show brightness modulations due to starspot activity, visualised in Fig.~\ref{fig:spot}. It can be seen that the variations are on a period consistent with the orbital period of the system, which in the previous section was interpreted as an indicator of rotational synchronisation. The variations evolve on a timescale of roughly two to three times the orbital period and have an amplitude of up to 0.006~mag.

\section*{Summary and conclusions}

\targ\ is dEB containing two late-F stars in a circular orbit with a period of 3.261~d. The most interesting features are total eclipses and a plethora of photometry from TESS, which together allow the radii of the stars to be obtained to very high precision. The light curves also show starspot modulation on the orbital period, indicating the stars are tidally synchronised. We see no evidence for pulsations, orbital eccentricity, or changes in the orbital period.

Two sets of RVs have been published for \targ, and they lead to slightly different measurements for the masses of the components. We are not aware of a reason to prefer one set over the other, so instead combined them to obtain spectroscopic orbits intermediate between the two sets. The uncertainties in $K_{\rm A}$ and $K_{\rm B}$ dominate those in both the mass and radius measurements. \gaia\ Data Release 4 (DR4\footnote{{\texttt https://www.cosmos.esa.int/web/gaia/data-release-4}}) is expected to provide extensive new RVs and thus a casting vote over which set of RVs to use (if either). 

The photometric properties of \targ\ are now extremely well-determined, although the spectroscopic orbits are not. New spectroscopic observations would be useful in determining the photospheric temperatures and chemical abundances of the component stars. The duration of totality is approximately 17 minutes, so a carefully-scheduled observation at secondary eclipse could record the spectrum of star~A without contamination by star~B.


\section*{Acknowledgements}

We thank the anonymous referee for a useful report which led to improvements in several parts of the paper.
This paper includes data collected by the TESS\ mission and obtained from the MAST data archive at the Space Telescope Science Institute (STScI). Funding for the TESS\ mission is provided by the NASA's Science Mission Directorate. STScI is operated by the Association of Universities for Research in Astronomy, Inc., under NASA contract NAS 5–26555.

This paper includes observations made with the Isaac Newton Telescope operated on the island of La Palma by the Isaac Newton Group of Telescopes in the Spanish Observatorio del Roque de los Muchachos of the Instituto de Astrof\'{\i}sica de Canarias.

This work has made use of data from the European Space Agency (ESA) mission {\it Gaia}\footnote{\texttt{https://www.cosmos.esa.int/gaia}}, processed by the {\it Gaia} Data Processing and Analysis Consortium (DPAC\footnote{\texttt{https://www.cosmos.esa.int/web/gaia/dpac/consortium}}). Funding for the DPAC has been provided by national institutions, in particular the institutions participating in the {\it Gaia} Multilateral Agreement.
The following resources were used in the course of this work: the NASA Astrophysics Data System; the SIMBAD database operated at CDS, Strasbourg, France; and the ar$\chi$iv scientific paper preprint service operated by Cornell University.



\end{document}